# Can apparent superluminal neutrino speeds be explained as a quantum weak measurement?

M V Berry[1], N Brunner[1], S Popescu[1] & P Shukla[2]

[1] H H Wills Physics Laboratory, Tyndall Avenue, Bristol BS8 1TL, UK
[2] Department of Physics, Indian Institute of Technology, Kharagpur, India

**Abstract**

Probably not.

If recent measurements [1] suggesting that neutrinos travel faster than light survive scrutiny, the question of their theoretical interpretation will arise. Here we discuss the possibility that the apparent superluminality is a quantum interference effect, that can be interpreted as a weak measurement [2-5]. Although the available numbers strongly indicate that this explanation is not correct, we consider the idea worth exploring and reporting – also because it might suggest interesting experiments, for example on electron neutrinos, about which relatively little is known. Similar suggestions, though not interpreted as a weak measurement [6, 7] or not accompanied by numerical estimates [6, 8], have been proposed independently.

The idea, following analogous theory and experiment [9] involving light in a birefringent optical fibre, is based on the fact that the vacuum is birefringent for neutrinos. We consider the initial choice of neutrino flavour as a preselected polarization state, together with a spatially localized initial wavepacket. Since a given flavour is a superposition of mass eigenstates, which travel at different speeds, the polarization state will change during propagation, evolving into a superposition of flavours. The detection procedure postselects a polarization state, and this distorts the wavepacket and can shift its centre of mass from that expected from the mean of the neutrino velocities corresponding to the different masses. This shift can be large enough to correspond to an apparent superluminal velocity (though not one that violates relativistic causality: it cannot be employed to send signals). Large shifts, corresponding to states arriving at the detector that are nearly orthogonal to the polarization being detected, are precisely of the type considered in weak measurement theory.

It seems that only muon and tau neutrino flavours are involved in the experiment, and we denote these by 1 (muon) and 2 (tau). The initial beam, with ultrarelativistic central momentum $\bar{p}$, is almost pure muon, which can be represented as a superposition, with mixing angle $\theta$, of mass states $|+\rangle$ and $|-\rangle$, with $m_+ > m_-$:

$$|\Psi_0(x)\rangle = (\cos\theta|+\rangle + \sin\theta|-\rangle)\exp\left(i\frac{\bar{p}x}{\hbar}\right)f(x). \tag{1}$$

Here $f(x)$ represents the envelope of the initial wavepacket, normalized and centred on $x=0$. For convenience, and with no effect on the final results (7) and (8), we take this as a gaussian of width $W$, that is

$$f(x) = \frac{1}{\pi^{1/4}\sqrt{W}}\exp\left(-\frac{x^2}{2W^2}\right). \tag{2}$$

The two mass states evolve with different phases and group velocities, so that the state arriving at the detector after travelling for time $t$ is

$$|\Psi(x,t)\rangle = \exp\left(i\frac{\bar{p}x}{\hbar}\right)\left[\cos\theta\exp\left(-i\frac{tE_+}{\hbar}\right)|+\rangle f(x - tv_+) \right.$$
$$\left. + \sin\theta\exp\left(-i\frac{tE_-}{\hbar}\right)|-\rangle f(x - tv_-)\right]. \tag{3}$$

This is an approximation, neglecting the spreading and distortion [10] of the individual packets – both negligible in the present case. $E_\pm$ and $v_\pm$ are the energies and group velocities of the two mass states, and we write

$$E_\pm = \bar{E} \pm \tfrac{1}{2}\Delta E, \quad v_\pm = \bar{v} \pm \tfrac{1}{2}\Delta v, \quad x = \bar{v}t + \xi, \tag{4}$$

in which the new coordinate $\xi$ measures deviation from the centre of the wavepacket expected by assuming it travels with the mean velocity. In the experiment, the detector postselects the muon flavour [1], so the final spatial wavepacket is

$$F(\xi,t) = N\exp\left(i\frac{\bar{p}x}{\hbar}\right)\left(\cos\theta\langle+| + \sin\theta\langle-|\right)|\Psi(\xi+\bar{v}t)\rangle$$

$$= N\exp\left(i\frac{(\bar{p}x - \bar{E}t)}{\hbar}\right)\left[\cos^2\theta\exp\left(-i\frac{t\Delta E}{2\hbar}\right)f\left(\xi - \tfrac{1}{2}t\Delta v\right)\right. \quad (5)$$

$$\left. + \sin^2\theta\exp\left(+i\frac{t\Delta E}{2\hbar}\right)f\left(\xi + \tfrac{1}{2}t\Delta v\right)\right],$$

where $N$ is a normalization constant.

Thus the shift in the measured final position of the wavepacket is found, after a short calculation, to be

$$\bar{\xi} = \int_{-\infty}^{\infty}d\xi\,\xi\left|F(\xi,t)^2\right| = \tfrac{1}{2}t\Delta v\,\frac{\cos 2\theta}{1 - \tfrac{1}{2}\sin^2 2\theta\left(1 - \cos\left(\frac{t\Delta E}{\hbar}\right)\exp\left(-\frac{(t\Delta v)^2}{4W^2}\right)\right)}. \quad (6)$$

In the prefactor, $t\Delta v$ is the relative shift of the two mass wavepackets, expected from the difference of their group velocities. This is multiplied by a factor representing the influence of the measurement – that is of the pre- and postselection and the evolution. If $t\Delta v$ is small compared with the width of the packet, as we will see that it certainly is in the neutrino case, the shift simplifies to

$$\bar{\xi} = \tfrac{1}{2}t\Delta v\,\frac{\cos 2\theta}{1 - \sin^2 2\theta \sin^2\left(\frac{t\Delta E}{2\hbar}\right)}, \quad (7)$$

which involves neither the width nor the shape of the wavepacket. When interpreted as an effective velocity shift, that is

$$\Delta v_{\text{eff}} = \frac{\bar{\xi}}{t} = \tfrac{1}{2} \Delta v \frac{\cos 2\theta}{1 - \sin^2 2\theta \sin^2\left(\frac{t\Delta E}{2\hbar}\right)}, \tag{8}$$

this is the same as the result found in [6].

As in the optical analogue [9], we can interpret the shift as a weak value in a measurement of the velocity difference operator

$$\Delta \hat{v} = \tfrac{1}{2} \Delta v \sigma_z = \tfrac{1}{2} \Delta v \begin{pmatrix} 1 & 0 \\ 0 & -1 \end{pmatrix}, \tag{9}$$

which as shown in [8] appears naturally from the Dirac equation for the neutrinos. The preselected state, after the evolution and before the measurement, is

$$|\text{pre}\rangle = \cos\theta \exp\left(-\mathrm{i}\frac{tE_+}{\hbar}\right)|+\rangle + \sin\theta \exp\left(-\mathrm{i}\frac{tE_-}{\hbar}\right)|-\rangle, \tag{10}$$

and the postselected state is

$$|\text{post}\rangle = \cos\theta|+\rangle + \sin\theta|-\rangle. \tag{11}$$

Standard weak measurement theory [3] now gives the weak value

$$\Delta v_{\text{weak}} = \frac{\langle \text{post}|\Delta \hat{v}|\text{pre}\rangle}{\langle \text{post}|\text{pre}\rangle} = \tfrac{1}{2}\Delta v \frac{\cos 2\theta \cos\frac{t\Delta E}{2\hbar} - \mathrm{i}\sin\frac{t\Delta E}{2\hbar}}{\cos\frac{t\Delta E}{2\hbar} - \mathrm{i}\cos 2\theta \sin\frac{t\Delta E}{2\hbar}}. \tag{12}$$

This is a complex quantity [11], whose real part is identical with the velocity shift (8):

$$\text{Re}\, \Delta v_{\text{weak}} = \Delta v_{\text{eff}}. \tag{13}$$

The possibility of superluminal velocity measurement arises because the amplification factor in (8) can be arbitrarily large if $\sin^2 2\theta$ and $\sin^2\left(\dfrac{t\Delta E}{2\hbar}\right)$ are close to unity, corresponding to near-orthogonality of $|\text{pre}\rangle$ and $|\text{post}\rangle$. An illustration of the distorted wavepacket $F(\xi,t)$, for a case where the amplification factor is large, is shown in figure 1 (for a more general and detailed discussion of wavefunctions ('pointer states') after postselection, see [12], especially section 6).

For neutrinos with momentum $p$, the energies of the two mass states, in the ultrarelativistic regime, are

$$E_\pm = \sqrt{p^2 c^2 + m_\pm^2 c^4} \approx pc + \frac{m_\pm^2 c^3}{2p}, \quad \text{i.e.} \quad \Delta E = \frac{(m_+^2 - m_-^2)c^3}{2p}, \tag{14}$$

and the group velocities are

$$v_\pm = \frac{\partial E_\pm}{\partial p} \approx c - \frac{m_\pm^2 c^3}{2p^2}, \quad \text{i.e.} \quad \Delta v = -c\frac{(m_+^2 c^4 - m_-^2 c^4)}{2(pc)^2} = -\frac{\Delta E}{p}. \tag{15}$$

Thus $\Delta v<0$, so, in order for the apparent velocity to be superluminal, $\Delta v_{\text{eff}}$ in (8) must be positive; this can be accommodated by making $\cos 2\theta$ negative.

Note also that $v_+$ and $v_-$ are less than $c$ if both neutrino masses are nonzero, so the individual mass eigenstate wavepackets move with subluminal group velocities; any superluminal velocity arising from (8) is a consequence of pulse distortion (illustrated in figure 1), associated with the postselection, i.e. considering only arriving muon neutrinos. In the more conventional superluminal wave scenario [10], group velocities faster than light, and the pulse distortions that enable them to occur, are associated with propagation of frequencies near resonance, for which there is absoption, i.e. nonunitary propagation. That is also true in the optical polarization experiments [9] and in the neutrino situation considered here, with the difference that the nonunitarity, which gives rise to the superluminal velocity, is not continuous during propagation but arises from the sudden projection onto the postselected state.

In the experiment, the energies of the neutrinos varied over a wide range, with an average of $cp = 28.1 \text{GeV}$. For the difference in the squared masses, with electron neutrinos neglected and $m_+$ and $m_-$ identified with the standard $m_2$ and $m_3$, a measured value [13] is $m_+^2 c^4 - m_-^2 c^4 \approx 2.43 \times 10^{-3} \text{eV}^2$. This gives

$$\frac{\Delta v}{c} = -1.5 \times 10^{-24}. \tag{16}$$

The apparent velocity measured in the experiment [1] was $(1 + 2.5 \times 10^{-5})c$. Comparison with the quantum velocity shift $\Delta v_{\text{eff}}$ in (8) would require knowlege of $m_+$ and $m_-$, not just their squared difference, and the individual masses are not known. But even on the most optimistic assumption, that $m_-=0$, it is immediately clear that it is unrealistic to imagine that the quantum amplification factor in (8) can bridge the gap of 19 orders of magnitude between (16) and the measured superluminal velocity. This pessimistic conclusion is reinforced by an explicit estimate. The amplification factor in (8) can exceed unity only if $\sin^2 \frac{t \Delta E}{2\hbar} > \frac{1}{2}$; but from the path length $d$=730km in the experiment, the accumulated phase is

$$\frac{t \Delta E}{2\hbar} = \frac{dE}{2c\hbar} \frac{\Delta v}{c} \approx 0.16 \approx 9.4° \text{ i.e. } \sin^2 \frac{t \Delta E}{2\hbar} \approx 0.03, \quad (17)$$

so there is no amplification at all. As further reinforcement of the conclusion that we have been discussing a small effect, measurements of the mixing angle [13] give $\sin^2\theta$>0.9, i.e. $|\cos\theta|$<0.32, so the numerator in (5) is small. Similar conclusions were reached by an analysis [7] that also considered the wide energy range of the neutrino beam. Finally, the estimate $t \Delta v = d \frac{\Delta v}{c} \approx 10^{-18}$ m confirms the approximation of (6) by (8) for any conceivable pulse width $W$.

**Figure caption**

Figure 1. Full curve: hypothetical neutrino wavefunction intensity $|F(\xi,t)|^2$ after postselection (equation 5); dotted curves: intensities of each component gaussian, whose coherent superposition gives $F(\xi,t)$. Parameters (chosen to illustrate distortion, not physical neutrinos) are: mixing angle $\theta=0.99\pi/4$, phase $t\Delta E/2\hbar = \frac{1}{2}\pi$ (so the two gaussians are in antiphase), separate shifts (thin vertical lines) $\frac{1}{2}t\Delta v = 1$, mean neutrino position (equation 6) (thick vertical line) $\overline{\xi} = 2.675$.

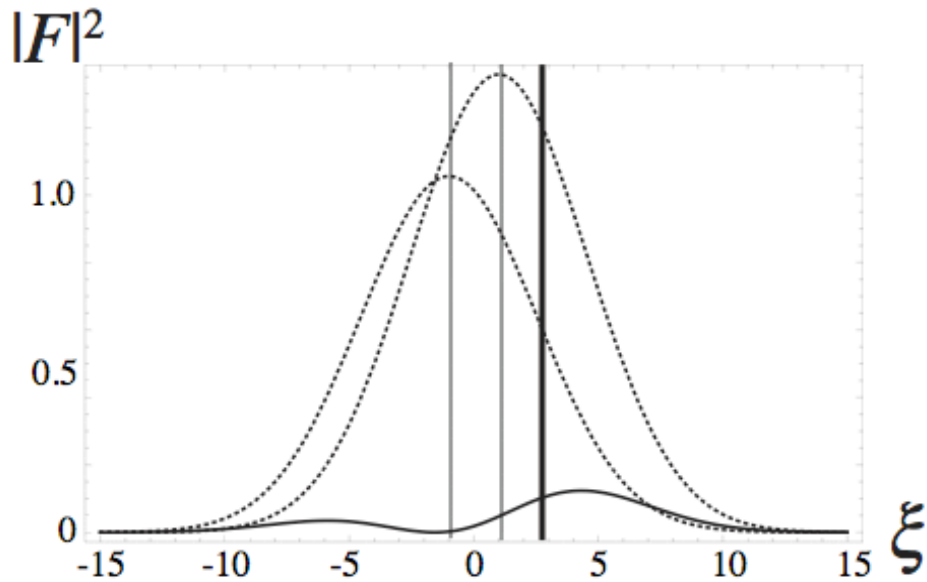

figure 1